 \documentclass[preprint2]{aastex}

\slugcomment{}

\usepackage{graphicx}
\usepackage{natbib}
\bibpunct{(}{)}{;}{a}{}{,}

\shorttitle{X-rays from HD 189733}
\shortauthors{Pillitteri et al.}

\begin{document}


\title{X-ray activity phased with planet motion in HD 189733? }


\author{I. Pillitteri\altaffilmark{1} 
\and H. M. G\"unther\altaffilmark{1} 
\and S. J. Wolk\altaffilmark{1}
\and V. Kashyap\altaffilmark{1} 
\and O. Cohen\altaffilmark{1} }
\affil{Harvard-Smithsonian Center for Astrophysics, 
60 Garden Street, Cambridge, MA 02138}

\begin{abstract}
We report on the follow-up XMM-Newton observation of { the planet-hosting star} HD~189733
we obtained in April 2011. We observe a flare just after the 
secondary transit of the hot Jupiter. This event shares the same phase and many of the
characteristics of the flare we observed in 2009. 
We suggest that a systematic interaction between planet and stellar magnetic 
fields when the planet passes close to active regions on the star can lead to periodic
variability phased with planetary motion.
By mean of high resolution X-ray spectroscopy with RGS we determine that the corona 
of this star is unusually dense.
\end{abstract}


\section{Introduction}
The evidence of star-planet interaction (SPI) in the X-ray band is a lively matter of debate.
To first order, close-in giant planets (also known as ``hot Jupiters'') 
should affect their host stars through both tidal and magneto-hydrodynamical effects 
(cf. \citealp{Cuntz2000} and \citealp{Ip2004}).  
Both effects should scale as the $-3$ power distance between the bodies (\citealp{Saar2004}).
\citet{Kashyap08} showed that stars with hot Jupiters are statistically 
brighter in X-rays than stars without hot Jupiters. 
On average \citet{Kashyap08} observed an excess of X-ray emission by a factor of 4
in the hot Jupiter sample.
Interplay between the magnetic fields of the hot Jupiter and the star may be the source 
of this difference. This could be due to interacting winds and magnetic fields 
or indirectly by enhancing the stellar dynamo.
On the other hand, \citet{Poppenhager2010} (and \citealp{Poppenhager2011})
have found no statistical evidence of X-ray SPI as claimed by \citet{Kashyap08}.

The system of HD~189733\ offers a unique environment to study SPI 
effects in X-rays and disentangle them from proper coronal activity. 
It is composed of a K1.5V type star (at only 19.3 pc from Sun), 
and a M4 companion at 3200 AU from the primary, orbiting on a plane perpendicular 
to the line of sight. 
It hosts a hot Jupiter class planet (HD~189733 b) at a distance of only 0.031 AU 
with a orbital period of $\sim2.22$d \citep{Bouchy05}.

In 2009 we observed the eclipse of the planetary companion to  HD~189733 with
the goal of studying star planet interactions (SPI) in the case of a hot Jupiter
(\citealp{Pillitteri2010}, hereafter Paper I).  
{ We observed a {\em softening} of the spectrum in  
strict correspondence of the planetary eclipse,} and a flare which occurred 
3~ks after the end of the eclipse of the planet. The non-detection of the M type companion 
is a strong constraint on the age of the system at $\ge 1.5-2$ Gyr.

The high age of the secondary is interesting because it is inconsistent with age of the system 
as derived from stellar activity which is of order 600 Myr \citep{Melo06}.
Recently,  \citet{Schroeter2011} reported on Chandra observations of an planetary transit 
of Corot 2A finding a similar case. 
While they do not detect the transit in X-rays they find that the primary is 
X-ray bright with a luminosity $\sim1.9\cdot 10^{29}$ erg s$^{-1}$, indicating an 
age $<$ 300 Myr.  
Meanwhile, a potential stellar companion was undetected down to a limit of $L_X \sim9\cdot10^{26}$
 ergs s$^{-1}$  which is inconsistent with the 300 Myr age and the distance of 270 pc.

The beginning of the flare observed in 2009 in HD~189733 is at phase $\phi \sim 0.54$, 
which coincides with a location { 77$^{\rm o}$ forward of the sub-planetary point and
emerging to the day side of the star.} 
This is almost exactly the location of the magnetic sub-planetary 
point as calculated by \citet{Lanza08}.
The flare could be associated with the emergence of the foot-point of the magnetic
column to the earth facing side or a complex active region induced by magnetic SPI.
Overall, the flare is associated with a change of the mean plasma temperature from 
$\sim0.5$ keV to $\sim0.8$ keV. 
The \ion{O}{7}  triplet is in excess with respect to the best fit model. In the RGS spectra
we observed that during the flare the inter-combination line seems to disappear, 
and the forbidden line is less luminous. 

In this letter we report on the follow up observation obtained in April 2011 
at the same phase as in 2009, during an eclipse of the planet.
Sections \ref{observation} { and \ref{results0}} describe the observation and the results. 
Sect. \ref{conclusions} reports our conclusions.

\section{Observation and data analysis} \label{observation}
The observational setup mimics our previous observation made on May 18th 2009. 
We obtained a X-ray observation with {\em XMM-Newton} around the eclipse of 
the planet of HD~189733 on April 30th 2011, starting at 23:14.20 (Obs Id: 0672390201),
and for a total duration of $\sim$39.1 ks. As in the previous observation, we used the {\em Medium} filter.
The time between observations of  HD~189733 was almost 2 years or exactly 61530.1 ks 
(mid-eclipse to mid-eclipse).   
The period of the star is 11.953$\pm$0.009 days \citep{Henry2008}, 
hence the star had rotated through 59.5795$\pm$0.045 periods, 
while the planet had orbited it 321 times. 

For the reduction of the data we followed the same procedure as in Paper I,
by using SAS ver. 11.0 to extract events, light curves and spectra of HD~189733\  
recorded with EPIC camera and RGS.  
For fitting the spectra, we used { 2-T} {\sc VAPEC} models for the pre-flare phases
(see description of light curve in Sect. \ref{results})
with different temperatures but coupled to have the same abundances. 
Abundances of Fe, Ne, and O were left free to vary while all other abundances 
were kept fixed at the solar values.
For flare and post-flare phases we added a third {\sc VAPEC} component keeping
frozen the parameters of the first two {\sc VAPEC} components. The abundances of the
third component were linked and fixed. We obtained estimates of the temperature 
and emission measure of the flaring plasma, also following its fading after the flare. 
We applied the same procedure of best fit to the spectra 
obtained in 2007 (for the whole observation), and 2009 (split in pre-flare, 
flare and post-flare phases) in order to compare the results.

We merged the RGS spectra in the non-flare and in the flare state of all observations. 
Flare and non-flare state were fitted independently using Gaussian 
lines with a narrow intrinsic line width in the SHERPA fitting tool 
\citep{2011ASPC..442..687R}, so that the total line width is dominated by 
instrumental broadening. Lines are adjusted in wavelength to account for 
possible errors in the zero-point of the wavelength calibration, 
but the wavelength difference in multiplets is hold constant. 
Within the errors, all wavelength are compatible with the theoretical values. 
Instrumental background and source continuum are assumed to be constant over 
a small wavelength region around the fitted lines.
{ The fits are done using a Cash statistic, which 
takes into account the Poisson distribution of counts, but the estimates 
of errors might be uncertain for very low count numbers.}

\begin{table*}
\caption{\label{fitpn} Values of best fit modeling to spectra before, during and after the
flare, respectively for 2011 and 2009 observations. We report also the fit of 2007 observation
with two 2-T {\sc VAPEC} components. The abundances of Fe, Ne and O in 2011 are derived from
the pre flare phase, with values: Fe = 0.57$_{-0.15}^{+0.11}$, O =  0.51$_{-0.07}^{+0.09}$, 
Ne = 0.3$_{-0.3}^{+0.07}$, respectively. { The values of emission measures are given in units of
$10^{50}$ cm$^{-3}$}.}
\begin{center}
\small
\begin{tabular}{c | c c c c c c c c}\hline\hline
Phase & kT$_1$  & kT$_2$ & kT$_3$ & E.M.$_1$  & E.M.$_2$ & E.M.$_3$ & log $f_X$ & log $L_X$  \\
       &  (keV)  &  (keV) & (keV)  & (cm$^{-3}$)   & (cm$^{-3}$)  &  (cm$^{-3}$) & (erg s$^{-1}$ cm$^{-2}$) & (erg s$^{-1}$) \\\hline
2011 Pre    & 0.24 $_{-0.03}^{+0.02}$ & 0.73 $_{-0.11}^{+0.08}$ & -- & 5.8 $_{-0.9}^{+1.0}$ & 3.6$_{-0.9}^{+0.9}$ & -- &  -12.5  & 28.17 \\ 
Flare & (0.24) & (0.73) & 0.9$_{-.1}^{+0.1}$ & (5.8) & (3.6) & 3.0$_{-0.5}^{+0.5}$ & -12.36 & 28.29 \\
Post  & (0.24) & (0.73) & 0.62$_{0.2}^{0.2}$ & (5.8) & (3.6) & 1.35$_{-0.03}^{+0.03}$ & -12.41 & 28.24 \\
\hline
2009 Pre & 0.18$_{-0.08}^{+0.08}$ & 0.47$_{0.08}^{+0.08}$ & -- & 4.1$_{-1.8}^{+1.8}$ & 5.6$_{-2.2}^{+2.3}$ & -- & -12.50 & 28.15 \\
Flare    & (0.18) & (0.47) & 0.99$_{-0.08}^{+0.08}$ & (4.1) & (5.6) & 3.2$_{-0.5}^{+0.4}$ & -12.37 & 28.29 \\
Post     & (0.18) & (0.47) & 0.75$_{-0.17}^{+0.17}$ & (4.1) & (5.6) & 1.3$_{-0.4}^{+0.3}$ &  -12.4 & 28.22 \\\hline
2007 & 0.24$_{-0.01}^{+0.01}$ & 0.71$_{0.03}^{+0.04}$ & -- & 4.7$_{-0.3}^{+0.4}$ & 2.8$_{-0.3}^{+0.3}$ & -- & -12.60 & 28.05 \\
\hline \hline
\end{tabular}
\end{center}
\end{table*}

\section{Results}
\label{results0}
\subsection{PN light curve and spectra}
\label{results} 
\begin{figure}
\begin{center}
\includegraphics[width=0.99\columnwidth]{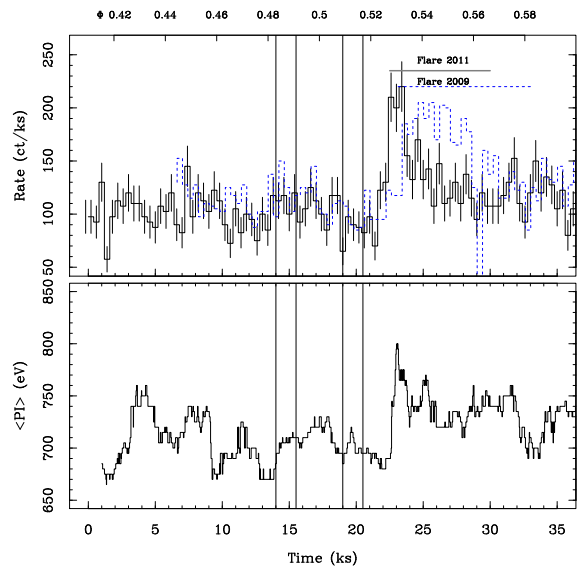}
\end{center}
\caption{\label{lcpn} Top panel: light curve of PN in 0.3 -- 1.5 keV.
The 2009 light curve (dotted line) is shown in phase with the 2011 light curve. 
Phase is marked on the top axis. The time bins are 400 sec.
Bottom panel: median of energy (PI) as a function of time. { The error bar
of $<PI>$ curve is about $12-15$ eV.}
Vertical lines mark 1st contact to 4th contact.
Light curves of $<\mathrm{PI}>$ are smoothed  
{ by taking the median of the sample of 200 events and varying the sample by adding 
five new events and removing the five oldest ones} (c.f. Paper I).}
\end{figure}

Fig. \ref{lcpn} shows the light curve of EPIC PN in 0.3--1.5 keV 
 \footnote{Ephemerides in {\sc http://var2.astro.cz/ETD/etd.php?STARNAME=\newline HD189733\&PLANET=b} 
 \citep{Poddany2010}}.
The overall rate in 2011 is similar to the PN rate recorded in 2009 and
about twice the PN rate in 2007 observation (quiescent rate in 2009 and 2011 
$\sim100\pm12$ ct ks$^{-1}$, quiescent rate in 2007: $\sim 60\pm 10$ ct ks$^{-1}$).
The average energy of the spectrum in 2011 before and during the eclipse is $675\pm20$ eV, 
it has been $700\pm 10$ eV in 2009 before the eclipse and $660\pm10$ eV during the eclipse.
During the planetary eclipse we do not see the softening as in 2009. 

The most striking feature is the flare after the end of the planetary eclipse, 
which is analogous to the main flare seen in the 2009 observation. In 2011, the flare 
starts at phase 0.52, while it starts at 0.54 in 2009. 
The duration of the flare ($\sim7$ ks, { evaluated by eye}),
the peak rate ($\sim230$ ct ks$^{-1}$ vs. $\sim210$ ks$^{-1}$ in 2009) and the presence 
of secondary impulses during the decay are quite similar to those observed in 2009 as well. 
{ The detection of two flares within  120 ks (i.e. the sum of exposure times in
2007, 2009 and 2011 observations) is interesting on its own, given that for active stars 
the typical rate of occurrence of bright flares is one every 500 ks 
(\citealp{Wolk2005}, and \citealp{Caramazza2007}).}

\begin{figure}
\begin{center}
\includegraphics[width=0.99\columnwidth]{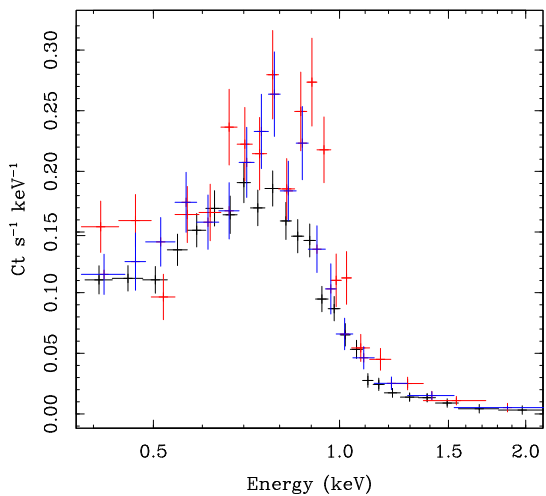}
\end{center}
\caption{\label{specpn} Spectra of PN before (black), during (red) and after the flare (blue). 
Differences in the shape are visible in the flare spectrum especially around 0.9 keV.}
\end{figure}

Table \ref{fitpn} reports the best fit values of the
spectra before, during and after this flare. For comparison, we report  also the
values obtained with the same scheme of best fit for 2007 and 2009 observations.
Fig. \ref{specpn} shows the PN spectra accumulated before, during, and after the bright
flare observed in 2011. 

The pre-flare phase has a best fit with two thermal components, 
at 0.24 keV and 0.73 keV, respectively. In 2009 the values are lower
(0.18 keV and 0.47 keV, respectively) 
and thus the corona appears colder than in 2011.

During the flare the plasma has a marginally higher temperature, around 0.9 keV
(flare temperature in 2009 was $\sim1$ keV).
In facts, the flare spectrum shows an excess around 0.9 keV (Fig. \ref{specpn}). 

In the post-flare phase the plasma cools essentially to the pre-flare temperature.
The third component has a temperature similar to the hot component of the pre-flare and it
is almost indistinguishable from the latter. It is worth noting that the luminosity
of the star remains slightly higher after the flare. This is observed in both 2011 and 2009.

As suggested in Paper I, this flare could arise from an active region 
which sits in a well defined location on the stellar surface. 
It is plausible that in this region a strong magnetic field is present  
given its configuration. 
The periodic passage of the planet could trigger magnetic reconnections and thus strong flares 
that could not arise elsewhere on the stellar surface. 
\citet{Fares2010} have published maps of the configuration of the magnetic field of HD~189733\ 
in 2006, 2007 and 2008. The field has a complex configuration with a toroidal component 
up to 40G in 2008 (the closest to the 2009 observation and the present one) 
and it changed configuration between 2006 and 2008. 
The dipolar component is strongly non-axisymmetric in 2008. 
\citet{Cohen2011} have shown through MHD simulations based on these maps that magnetic SPI is 
possible whenever the planetary and stellar alfv\`enic surfaces intersect each other. 
Given the non-symmetric shape of the stellar magnetic field, 
the interaction of planetary and stellar fields are not occurring continuously 
but at some phase determined by the planetary motion and the stellar rotation.

The emission levels in X-rays in 2009 and 2011 are quite similar.
On the other hand, these are $\sim80\%$ greater than in 2007. 
The observations suggests a cyclic activity which is in agreement with the changes 
in the magnetic field  derived by \citet{Fares2010}.
We have no information on the topology of the magnetic field during last X-ray observation.
The hypothesis that periodic occurrence of strong variability is triggered by the planetary motion
needs further observations to be assessed. 


\subsection{RGS spectroscopy}
Merging the RGS spectra of two observations of {\em XMM-Newton} at the secondary eclipse and the one
at the planetary transit (2007, PI: Wheatley), we have obtained the first high resolution
spectroscopy { in X-rays} of HD~189733.
Table \ref{rgslines} shows line strengths of \ion{Fe}{17} and 
the He-like \ion{Ne}{9}, \ion{O}{7} and the H-like \ion{O}{8}. 
This latter is the only line which is significantly detected 
by both RGS detectors, the line fluxes are compatible within the statistical errors.
{ Due to CCD failures, \ion{O}{7} is visible only in RGS~1, while \ion{Ne}{9} is visible
only in RGS~2.}

\begin{table*}
\caption{Lines with significant detections. Error ranges are $1\sigma$ or missing 
if unconstrained. \label{rgslines}}
\begin{center}
\begin{tabular}{cccccc}
\hline \hline
line & $\lambda$ & RGS1 & RGS1-flare & RGS2 & RGS2-flare \\
 & [\AA{}] & \multicolumn{4}{c}{$10^{-15}$ erg s$^{-1}$ cm$^{-2}$}\\ \hline
Ne X Ly$\alpha$ & 12.13 & ... & ... & $ 4.8_{-2.0}^{+2.2}$ & $14.1_{-7.5}^{+9.1}$ \\
Ne IX r & 13.45 & ... & ... & $ 7.4_{-2.6}^{+2.8}$ & $ 4.3_{-4.3}^{+6.5}$ \\
Ne IX i & 13.55 & ... & ... & $ 0.0_{...}^{+1.9}$ & $ 9.3_{-4.5}^{+7.4}$ \\
Ne IX f & 13.7 & ... & ... & $ 8.2_{-2.7}^{+3.1}$ & $ 0.0_{...}^{+21.6}$ \\
Fe XVII & 16.78 & $ 9.7_{-2.3}^{+2.5}$ & $10.9_{-6.1}^{+7.4}$ & ... & ... \\
O VIII Ly$\alpha$ & 18.97 & $19.3_{-3.0}^{+3.1}$ & $21.9_{-8.0}^{+9.4}$ & $18.5_{-2.9}^{+3.2}$ & $39.2_{-8.7}^{+9.5}$ \\
O VII r & 21.6 & $12.3_{-2.2}^{+2.5}$ & $15.2_{-6.7}^{+8.8}$ & ... & ... \\
O VII i & 21.81 & $ 8.4_{-1.9}^{+2.2}$ & $ 0.0_{...}^{+5.4}$ & ... & ... \\
O VII f & 22.1 & $ 8.9_{-2.0}^{+2.2}$ & $ 8.0_{-5.3}^{+7.4}$ & ... & ... \\
\hline
\end{tabular}
\end{center}
\end{table*}

\subsubsection{Temperature}
In the flares, the \ion{Ne}{10} emission increases by about a factor of three in agreement with  
the increased temperature seen in the light curve of median energy and spectra of PN. 
The total emission of the He-like triplets is similar in flare and non-flare intervals.

\begin{figure}
\includegraphics[width=0.99\columnwidth]{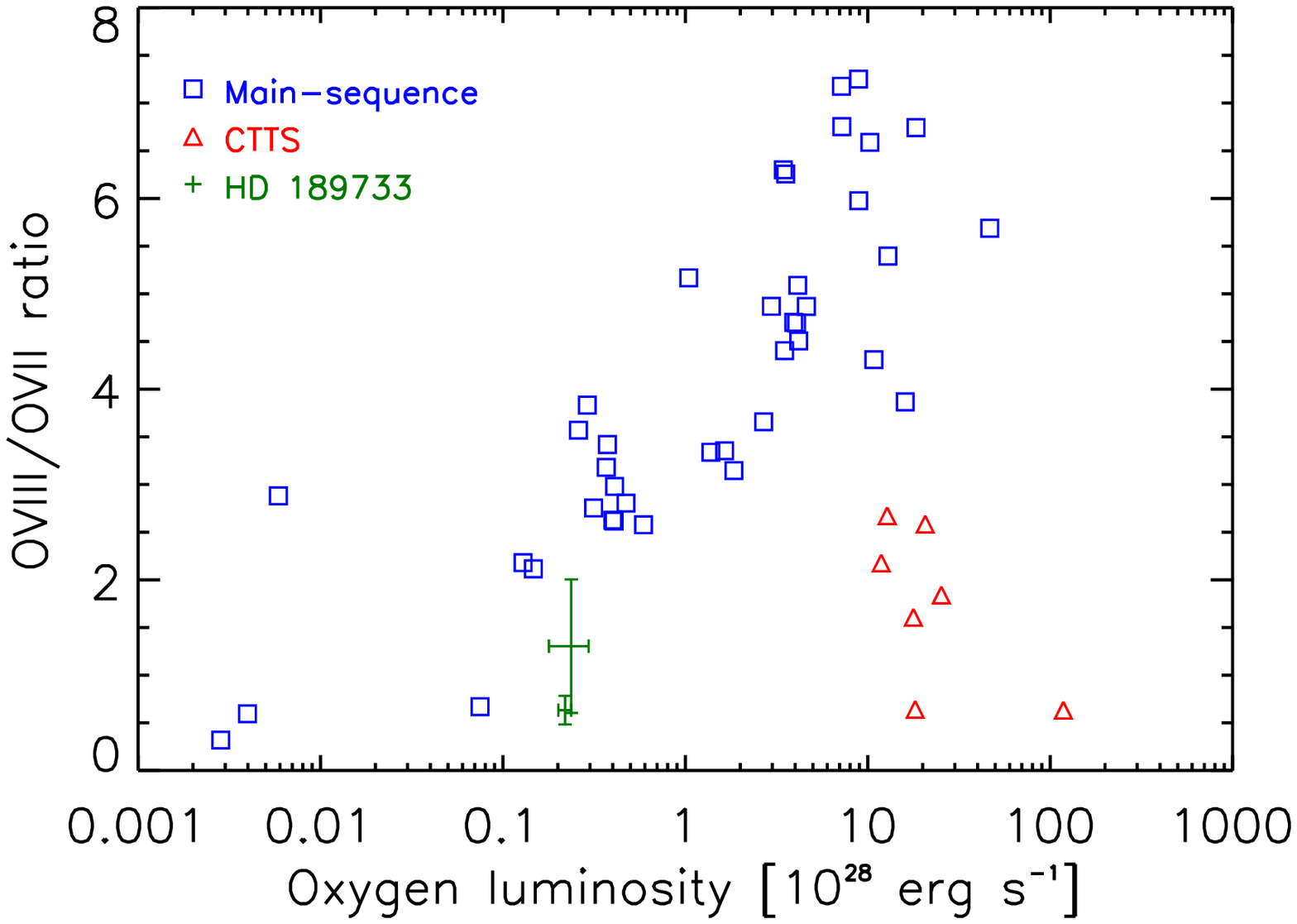}
\caption{Ratio of \ion{O}{8}/\ion{O}{7} emission compared to the total oxygen luminosity. 
The measurements for HD~189733 are shown with errors. The flare measure has
larger error bars because of the lower count number. See text for details (color in electronic version).
\label{fig:o82o7} }
\end{figure}

The ratio of lines from different ionization stages can be used as a temperature diagnostic 
\citep{Mewe1991}. 
Figure~\ref{fig:o82o7} shows the ratio of \ion{O}{8}/\ion{O}{7} emission compared to the total 
luminosity in these lines. Blues squares show main-sequence (MS) stars from the sample of 
\citet{2004A&A...427..667N}, red triangles are classical T Tauri stars (CTTs), which are accreting
pre-main sequence stars \citep{RULup,2011AN....332..448G}. CTTs show an excess of cool emission, 
which is likely powered by accretion shocks. 
HD~189733 is among the stars in the sample with the lowest X-ray luminosity and 
\ion{O}{8}/\ion{O}{7} ratio. Thus, it is marginally cooler than MS stars of comparable 
luminosity, { but its corona is over-dense like PMS accreting stars}.
All CTTs in Fig.~\ref{fig:o82o7} are significantly brighter than HD~189773, but this is a 
selection bias because only the brightest CTTs can be observed with X-ray gratings.

\subsubsection{Density}
The ratio of the forbidden line $f$ and the inter-combination line $i$ in the He-like triplets is 
density sensitive \citep{Mewe1991,2000A&AS..143..495P}. In the low-density, limit $f/i$ is 3.5 for 
\ion{O}{7} ($\log n_e < 9$) and 3.2 for \ion{Ne}{9} ($\log n_e < 10$) 
according to the CHIANTI database \citep{CHIANTI,2009A&A...498..915D}. 
We performed Monte-Carlo simulations to account for the non-gaussian errors in the distribution 
\citep[][appendix A]{HD163296}. The \ion{Ne}{9} lines are compatible with the low to moderate density 
limit in the non-flare state (95\% confidence limit $f/i > 2.2$, i.e. $\log n_e < 11$), 
in the flare spectrum the error is too large to constrain the density.
The non-detection of the \ion{O}{7} $i$ line in the flare is consistent with moderate densities, 
the 95\% confidence lower limit is $f/i=0.5$ ($\log n_e < 11.2$). 
In the non-flare state, the 95\% confidence upper and lower boundaries on the $f/i$ ratio are 1.8 
($\log n_e > 10.5$) and 0.6 ($\log n_e < 11.1$). 
Generally, the coronal emission in MS stars is in the low-density limit 
\citep{2004A&A...427..667N}, only in a few bright flares higher densities have been seen 
\citep[e.g. on Proxima Cen,][]{2002ApJ...580L..73G,2004A&A...416..713G}.

The total luminosity in the He-like triplets changes little between the flare and the non-flare 
state. This could indicate that they originate in a dense region which is not affected 
by flare heating. However, the tentative change in the $f/i$ ratio can be interpreted as a 
change in density, where the hotter plasma seen in the flare state has a lower density than in 
the non-flare state.

\section{Conclusions}
\label{conclusions}
We have analyzed a {\em XMM-Newton} observation of HD~189733\ at the secondary transit of the planet. 
We observed a flare with characteristics very similar to that observed
in the previous observation at the same phase. 
The recurrence of such flares is explained by the following scenario: 
an active region is present at the same location on the stellar 
surface in both observations. The magnetic interaction with the planet is inducing
a flaring activity in this region. The occurrence depends on the configuration of 
the stellar magnetic field and its strength. Further observations at the same phase can prove 
or reject this hypothesis. 

High resolution spectroscopy in quiescent and flaring state shows that the corona of 
HD~189733 is marginally cooler than MS stars with same oxygen line luminosity. 
We find  a marginal change in $f/i$ ratio of \ion{O}{7} lines in the flare state,
implying a lower density during flares with respect to quiescent state.
In summary, the corona of HD~189733 stands apart with respect to other coronae 
of Main Sequence stars without hot Jupiters. 
It is cold like the solar corona but it is dense and ten times more luminous and active. 
This could be related to SPI effects but more observations are needed to clarify 
the influence of its hot Jupiter on the corona.
\acknowledgments
The {\em XMM-Newton} guest investigator program supported IP and HMG through grant
NNX09AP46G. SJW was supported by NASA contract NAS8-03060 to the Chandra
Science Center. VK acknowledges CXC NASA contract NAS8-39073, and OC
acknowledges NASA LWSTRT grant NNG05GM44G.

{\it Facilities:} \facility{XMM-Newton}


\end{document}